\documentclass[12pt]{article}
\pdfoutput =1
\usepackage{float} %Include figure filesusepackage{graphicx} %Include figure files
\usepackage{amsmath}
\usepackage{slashed}
\usepackage{amsfonts}   
\usepackage{amssymb}

\begin{document}
\date{}
%%%%%%%%%%%%%%%%%%%%
\title{{\bf{\Large On pp wave limit for  $ \eta $  deformed superstrings}}}
%%%%%%%%%%%%%%%%%%%%
\author{
 {\bf {\normalsize Dibakar Roychowdhury}$
$\thanks{E-mail:  dibakarphys@gmail.com, Dibakar.RoyChowdhury@swansea.ac.uk}}\\
 {\normalsize  Department of Physics, Swansea University,}\\
  {\normalsize Singleton Park, Swansea SA2 8PP, United Kingdom}
\\[0.3cm]
}
%\date{}

\maketitle
%%%%%%%%%%%%%%%%%%%%%%%%%%%%%%%%%%%%%%%%%%%%%%%%%%%%%%%%%%%%%%%%%%%%%%%%%%%%
\begin{abstract}
In this paper, based on the notion of plane wave string/gauge theory duality, we explore the pp wave limit associated with the bosonic sector of $ \eta $ deformed superstrings propagating in $ (AdS_5 \times S^{5})_{\eta} $. Our analysis reveals that in the presence of NS-NS and RR fluxes, the pp wave limit associated to full ABF background satisfies type IIB equations in its standard form. However, the beta functions as well as the string Hamiltonian start receiving non trivial curvature corrections as one starts probing beyond pp wave limit which thereby takes solutions away from the standard type IIB form. Furthermore, using \textit{uniform} gauge, we also explore the BMN dynamics associated with short strings and compute the corresponding Hamiltonian density. Finally, we explore the Penrose limit associated with the HT background and compute the corresponding stringy spectrum for the bosonic sector. 
\end{abstract}
%%%%%%%%%%%%%%%%%%%%%%%%%%%%%%%%%%%%
\section{Overview and Motivation}
Two major obstructions towards understanding the celebrated AdS/CFT duality was the difficulty in the computation of anomalous dimensions associated with non BPS operators in the regime of strong coupling and the quantization of strings in $ AdS_5 \times S^{5} $ in the presence of RR five forms. However, there exists a miraculous way out to this problem namely, taking a specific limit \cite{Berenstein:2002jq} on both sides of the duality that simplifies the problem enormously. On the string theory side, this limit is achieved by considering strings that are boosted with the light like momentum along one of the isometry directions of $ S^{5} $ and thereby taking the so called Penrose limit around this null trajectory \cite{Berenstein:2002jq}-\cite{Itsios:2017nou}. The resulting geometry turns out to be manifestly invariant under $ SO(4) \times SO(4) $ symmetry of eight transverse directions together with the isometry along the time and the other angular direction. The last two isometries correspond to conserved charges like the energy ($ \mathfrak{w}_{n} $) and the angular momentum ($ J $) associated with the string states. On the gauge theory side, on the other hand, the large value of $ J $ corresponds to large $ R $- charge associated with single trace operators of the form, $ \mathcal{O}\sim tr[Z^{J}] $ where $ Z $ is some (complex) composite operator associated with the dual field theory \cite{Berenstein:2002jq}. Typically, the claim of the AdS/CFT duality is that at least for the planar limit, the energy associated with stringy excitations corresponding to a large value of $ J $ should be in one to one correspondence to that with the anomalous dimension associated with heavy nearly BPS like operators $ \mathcal{O} $ in the dual gauge theory. In other words, the spectrum on the both sides of the duality should match.

The purpose of the present article is to address the above mentioned issues in the context of $ \eta $ deformed superstrings \cite{Delduc:2013qra}-\cite{Sakamoto:2017wor} discovered in the recent years. The $ \eta $ deformation could be interpreated as a consistent way of $ q $ ($ \in \mathbb{R} $)-deforming the $ S $ matrix corresponding to $ AdS_{5}\times S^{5} $ superstrings. At the level of the classical string sigma model, the deformation parameter is introduced in the form of the so called Yang-Baxter (YB) deformations where the classical R matrix satisfies the modified YB equation. The question that we are interested here is whether the target space corresponding to $ \eta $ deformed supercosets could have any AdS/CFT like interpretations in terms of some (local) observables associated with quantum field theories in general. However, for the most purposes of the analysis, we stick to the stringy sector of the duality.

We start our analysis, in Section 2, by formally constructing the pp wave limit corresponding to $ \eta $ deformed $ AdS_5 \times S^{5} $ superstrings.
Our analysis reveals that the pp wave limit associated with the bosonic sector of the full superstring sigma model falls into one to one correspondence to that with the corresponding pp wave limit of $ AdS_5 \times S^{5} $ superstrings \cite{Berenstein:2002jq} (in the presence of the RR five form ($ \mathcal{F}_{5} $) couplings) except for a trivial rescaling of the mass of the world-sheet scalars,
\begin{eqnarray}
\hat{\mu}=\mu \frac{(1+\eta^{2})}{(1-\eta^{2})}
\end{eqnarray}
that eventually modifies the spectrum,
\begin{eqnarray}
\mathfrak{w}_{n}=\hat{\mu}\sqrt{1+ \frac{\lambda^{(\eta)}_{D} n^{2}}{J^{2}}};~~\lambda^{(\eta)}_{D}=\hat{\lambda} \frac{(1-\eta^{2})^{4}}{(1+\eta^{2})^{4}}
\end{eqnarray}
in the limit, $ J \rightarrow \infty $. Here, $ \hat{\lambda} $ is the modified t'Hooft coupling in the theory with $ \eta $ deformations \cite{Hoare:2014pna}. It turns out that the Killing vector field  \cite{Arutyunov:2015mqj} modifying the standard type IIB equations eventually becomes a constant and its effect on the corresponding beta functions is simply washed away in the pp wave limit. However, the Killing vector field starts receiving non trivial corrections beyond the pp wave limit which leads towards a deviation from the standard type IIB structure. Next, we go beyond the pp wave limit and compute the leading $ 1/J $ corrections to the Hamiltonian associated with the bosonic sector of the string sigma model.

In Section 3, we work out the Hamiltonian formulation for $ \eta $ deformed superstrings using the notion of uniform gauge \cite{Kruczenski:2004kw}-\cite{Arutyunov:2004yx} and in particular obtain the corresponding plane wave Hamiltonian expressed in terms of the effective coupling ($ \hat{\lambda} $) and the angular momentum ($ J $). In other words, we exploit the gauge freedom of the associated sigma model in order to set the world-sheet time ($ \tau $) equal to that of the target space time ($ t $) and one of the canonical momenta to that with the R symmetry generator ($ J $) in the dual field theory \cite{Arutyunov:2004yx}. Our analysis reveals that for short fluctuating strings propagating along the null geodesic on $ S^{5} $, one indeed obtains the $ \eta $ corrected (bosonic) pp wave Hamiltonian that depends on the perturbative coupling, $\frac{\hat{\lambda}}{J^{2}} $ and reduces correctly to that of the Hamiltonian corresponding to $ AdS_5 \times S^{5} $ superstrings \cite{Arutyunov:2004yx} in the limit of the vanishing deformation.  

In Section 4, we explore the pp wave limit of HT background \cite{Hoare:2015wia} that is related to the ABF background \cite{Arutyunov:2013ega} via T duality transformations.  The HT background is special in the sense that it satisfies the standard type IIB equations and the corresponding world-sheet theory is therefore Weyl invariant. However, our analysis reveals that unlike the case with the $ \eta $ deformations, four out of the eight scalar d.o.f receive mass in the pp wave limit of the HT background and the resulting equations of motion are quite non trivial in general. However, the dynamics gets simplified at a scale, $ L \gg \mu^{-1} $ and all the scalar modes essentially get decoupled and the theory could be solved in principle.

Finally, we draw our conclusion in Section 5.
%%%%%%%%%%%%%%%%%%%%%%%%%%%%%%%%%%%%%%%
\section{Deformations of $ AdS_5 \times S^{5} $}
The purpose of this Section is to discuss the spectrum associated with the bosonic sector of the full $ \eta $ deformed $ AdS_5 \times S^{5} $ superstring 
sigma model on the plane wave background. We start our analysis with a formal discussion on the generalized type IIB solutions \cite{Arutyunov:2015mqj} and then move over towards the rigorous construction of the pp wave limit following the systematic methodology developed for type IIB superstrings \cite{Berenstein:2002jq} in order to compute the corresponding stringy spectrum associated with the bosonic sector. We leave the corresponding analysis for fermions for future purposes.
%%%%%%%%%%%%%%%%%%%%%%%%%%%%%%%%%%%%%%%%%%%%
\subsection{The supergravity background}
Finding a standard type IIB solution corresponding to $ \eta $ deformed $ AdS_5 \times S^{5} $ string sigma model had remained as a puzzle since its very beginning. In \cite{Hoare:2015wia}, the authors had shown that T dualizing the six isometric (bosonic) directions of $ \eta $ deformed $ AdS_5 \times S^{5} $ solution one might formally end up in a type IIB solution which might be regarded as being that of the one parameter deformation of the standard $ AdS_5 \times S^{5} $ solution obtained via T duality. However, the resulting dilaton contains terms linear in isometric coordinates that precludes one to T dualize back into the original $ \eta $ deformed solution. Finally, as a non trivial check, one can show that in the limit of the vanishing deformation it is indeed possible to T dualize back to the original $ AdS_5 \times S^{5} $ solution.

Although $ \eta $ deformation preserves the fermionic kappa symmetry for the worldsheet theory, the corresponding target space fails to satisfy the standard type IIB equations \cite{Arutyunov:2015mqj}. It was pointed out in \cite{Arutyunov:2015mqj} that the kappa invariance of the $ \eta $ model imposes constraints on the 10d target space which leads to the \textit{generalized} type IIB supergravity equations (corresponding to NS-NS fields) of the following form\footnote{In the presence of RR fields these generalised equations (\ref{G1}) receive additonal $ \mathcal{F}^{2} $ contributions together with a set of second order equations corresponding to RR fields namely, $ \mathcal{F}=e^{\Phi}F $ \cite{Arutyunov:2015mqj}.} \cite{Arutyunov:2015mqj},
\begin{eqnarray}
R_{ab}-\frac{1}{4}H_{acd}H_{b}^{cd}= -D_{a}X_{b}-D_{b}X_{a}\nonumber\\
\frac{1}{2}D^{k}H_{kab}=X^{k}H_{kab}+D_{a}X_{b}-D_{b}X_{a}\nonumber\\
R -\frac{1}{12}H^{2}_{abc}+4 D_{a}X^{a}-4 X^{a}X_{a}=0\label{G1}
\end{eqnarray}
that follows from the scale invariance of the superstring sigma model. Here, the vector $ X_{a}=I_{a}+Z_{a} $ is intimately related to the associated Weyl invariance of the wroldsheet theory that satisfies the following equations \cite{Sakamoto:2017wor},
\begin{eqnarray}
D_{a}I_{b}+D_{b}I_{a}=0,~~I^{k}H_{kab}+D_{a}Z_{b}-D_{b}Z_{a}=0,~~I^{a}Z_{a}=0.
\end{eqnarray}
The dilation of the standard type IIB solution is included in,
\begin{eqnarray}
Z_{a}=\partial_{a}\Phi +U_{a}
\end{eqnarray}
which clearly reproduces the standard type IIB equations provided, $ I_{a}=0 $ and $ U_{a}=0 $.
The Weyl invariance associated with the $ \eta $ deformed worldsheet theory yields a \textit{constant} one loop beta function for the dilaton which is supposed to be zero in a critical string theory. In the subsequent analysis of this paper, we show a nice way out to all of these issues by taking the so called pp wave limit of the $ \eta $ deformed string sigma model that eventually turns out to be a critical string theory defined over a type IIB background.

%However, in a recent analysis \cite{Sakamoto:2017wor}, it was confirmed that for the NS-NS sector of the bosonic string theory defined on generalized gravity background, one can in principle write down the modified Fradkin–Tseytlin \textit{non-local} counter term as a one loop quantum correction that eventually restores the Weyl invariance in the so called double field theory framework which thereby ensures that the generalized supergravity equations fall under the same footing as those of the ordinary type IIB equations in a critical string theory. Restoration of the Weyl invariance is also quite suggestive from the fact that the generalized type IIB background (namely the $ \eta $ deformed target space) is related to the standard type IIB background through T duality transformations. Since T duality is a symmetry of the string theory therefore this suggests that the Weyl invariance has to be restored for the $ \eta $ model in some sense.

%%%%%%%%%%%%%%%%%%%%%%%%%%%%%%%%%%%%%%%%%%%%%%%%%%
\subsection{The pp wave background}
%%%%%%%%%%%%%%%%%%%%%%%%%%%%%%%%%%%%%%%
\subsubsection{Formal construction}
The purpose of this Section is to explore the Penrose limit corresponding to $ \eta $ deformed $ AdS_5 \times  S^{5}$ superstring sigma model following the standard prescription as in the case for the usual $ AdS_5 \times S^{5} $ superstrings \cite{Berenstein:2002jq}. In this part of our analysis, we focus only on the NS-NS sector of the full superstring spectrum and consider stringy dynamics over $ (AdS_5 \times S^{5})_{\eta} $ with NS fluxes. The background that is relevant for our analysis could be formally expressed in the so called global coordinates as\footnote{Notice that the parameter $ \kappa $ is related to the original deformation parameter $ \eta $ as, $ \kappa=\frac{2 \eta}{1-\eta^{2}} $ \cite{Arutyunov:2013ega}. Henceforth we will denote our deformation parameter as $ \kappa $.} \cite{Arutyunov:2013ega},
\begin{eqnarray}
ds^{2}_{ (AdS_{5})_{\kappa}}=L^{2}\left( -\frac{1+\varrho^2}{1-\kappa^2 \varrho^2}dt^2+\frac{d\varrho^{2}}{(1+\varrho^{2})(1- \kappa^{2}\varrho^{2})}+\frac{\varrho^{2}d\zeta^{2}}{1+\kappa^{2}\varrho^{4}\sin^{2}\zeta}\right)  \nonumber\\
+\frac{L^{2}\varrho^{2}\cos^{2}\zeta}{1+\kappa^{2}\varrho^{4}\sin^{2}\zeta} d\psi^{2}_{1}+L^{2}\varrho^{2}\sin^{2}\zeta d\psi^{2}_{2}\nonumber\\
ds^{2}_{ (S^{5})_{\kappa}}=L^{2}\left( \frac{1-r^2}{1+\kappa^2 r^2}d\psi^2
	+\frac{dr^2}{ \left(1-r^2\right)(1+\kappa^2 r^2)}+r^2 \sin^2\xi\, d\phi_2^2\right)  \nonumber\\ + \frac{L^{2}r^2}{1+\kappa^2r^4\sin^2\xi}\left( d\xi^2+\cos ^2\xi \, d\phi_1^2\right) \nonumber\\
B_{2AdS_5}=\frac{L^{2}\kappa}{2}\left( \frac{\varrho^{4}\sin 2\zeta}{1+\kappa^{2}\varrho^{4}\sin^{2}\zeta}d\psi_{1}\wedge d\zeta +\frac{2\varrho}{1-\kappa^{2}\varrho^{2}}dt \wedge d\varrho\right) \nonumber\\
B_{2S^{5}} =-\frac{\kappa L^{2}}{2} \left( \frac{r^4 \sin 2\xi}{1+\kappa^2 r^4\sin^2 \xi}d\phi_1\wedge d\xi + \frac{2r}{1+\kappa^2 r^2}d\psi\wedge dr\right).
\label{e19}
\end{eqnarray}
One should notice that the last two terms in the expansion for the $ B $ field (\ref{e19}) are the total derivative terms. These terms are therefore not important as far as the dynamics is concerned and one might therefore drop their contributions in the various subsequent expressions as well.

In order to proceed further, we set,
\begin{eqnarray}
\varrho =0,~r =\sin \theta
\end{eqnarray}
which yields the following metric together with the background $ B $ field,
\begin{eqnarray}
ds^{2}_{(R \times S^{5})_{\kappa}}=L^{2}\left( -dt^2+\frac{\cos^2 \theta}{1+\kappa^2 \sin^2 \theta}d\psi^2
	+\frac{d\theta^2}{ 1+\kappa^2 \sin^2\theta}\right)  \nonumber\\ + \frac{L^{2}\sin^2\theta}{1+\kappa^2 \sin^4\theta \sin^2\xi}\left( d\xi^2+\cos ^2\xi \, d\phi_1^2\right) +L^{2}\sin^2\theta \sin^2\xi\, d\phi_2^2\nonumber\\
B_{2S^{5}} =-\frac{L^{2}\kappa}{2} \left( \frac{\sin^4\theta \sin (2\xi)}{1+\kappa^2 \sin^4\theta \sin^2 \xi}d\phi_1\wedge d\xi + \frac{2 \sin \theta \cos \theta}{1+\kappa^2 \sin^2\theta}d\psi\wedge d\theta \right).
\end{eqnarray}

Our next task would be to explore light like geodesics parametrized as,
\begin{eqnarray}
t=t(\varsigma),~~\psi =\psi (\varsigma),~~\theta =0,~~\varrho =0.\label{E4}
\end{eqnarray}

Using (\ref{E4}), the action corresponding to the null geodesic associated with the relativistic point particle turns out to be,
\begin{eqnarray}
S=\frac{L^{2}}{2}\int d\varsigma e^{-1}(-\dot{t}^{2}+\dot{\psi}^{2}).
\end{eqnarray}

In order to explore the trajectory in the light cone frame, we introduce the following set of coordinates,
\begin{eqnarray}
x^{\pm}=\frac{1}{2}(t \pm \psi)\label{E6}
\end{eqnarray} 
which yields the following metric in the light cone frame,
\begin{eqnarray}
ds^{2}=-4 L^{2}dx^{+}dx^{-}.\label{E7}
\end{eqnarray}

Using (\ref{E7}), the equations of motion could be formally expressed as,
\begin{eqnarray}
\ddot{x}^{\pm}=0
\end{eqnarray}
which is trivially solved for the choice,
\begin{eqnarray}
x^{+}=C,~~x^{-}=\varsigma .
\end{eqnarray}

In order to explore the geometry near this null trajectory, we introduce following coordinates \cite{Plefka:2003nb},
\begin{eqnarray}
X^{+}=\frac{x^{+}}{\mu},~~X^{-}=\mu L^{2}x^{-},~~\varrho = \frac{R}{L},~~\theta =\frac{\Theta}{L}\label{E10}
\end{eqnarray}
and set the limit, $ L \rightarrow \infty $. Here, $ \mu $ is some mass scale introduced into the theory in order to maintain proper canonical dimension
for the coordinates.

Using (\ref{E6}) and (\ref{E10}), we can finally express the background metric in the following form,
\begin{eqnarray*}
ds^{2}=-4 dX^{+}dX^{-}- \hat{\mu}^2 (\Theta ^2 +R^{2})(dX^{+})^2+dR^{2}+R^{2}d\Omega^{2}_{3}\nonumber\\
+d\Theta^{2}+\Theta^{2}d\tilde{\Omega}^{2}_{3}+\frac{1}{L^{2}}(\Delta G)^{2}+\alpha^{2}(\Delta S)^{2}+\mathcal{O}(1/L^{4})\nonumber\\
\end{eqnarray*}
\begin{eqnarray}
(\Delta G)^{2}&=&\frac{\Theta^{4}}{3}(\mu^{2}(d X^{+})^{2}-d\tilde{\Omega}^{2}_{3})-R^{2}dR^{2}-2(R^{2}-\Theta^{2})dX^{+}dX^{-}\nonumber\\
d\Omega^{2}_{3}&=&d\zeta^{2}+\cos^{2}\zeta d\psi^{2}_{1}+\sin^{2}\zeta d\psi^{2}_{2}\nonumber\\
d\tilde{\Omega}^{2}_{3}&=&d\xi^{2}+\cos^{2}\xi d\phi^{2}_{1}+\sin^{2}\xi d\phi^{2}_{2}
\label{E11}
\end{eqnarray}
where we identify,
\begin{eqnarray}
\hat{\mu}=\mu \sqrt{1+\kappa^{2}}=\mu \frac{(1+\eta^{2})}{(1-\eta^{2})}
\end{eqnarray}
as being the rescaled mass parameter of the theory.

However, the significant contribution appears in the last term in (\ref{E11}) where we identify the following entity\footnote{One could think of the pp wave limit for the $ \kappa $ deformed background as an expansion in the parameter $ \alpha $ such that $ | \alpha L|\ll 1 $ for \textit{finite} background deformation $ \kappa $.}
\begin{eqnarray}
\alpha^{2}=\frac{\kappa^{2}}{L^{2}}
\end{eqnarray}
as being the additional dimensionful parameter of the theory that eventually contributes a term
\begin{eqnarray}
(\Delta S)^{2}=R^{2}dR^{2}-\Theta^{2}d\Theta^{2}-\mu^{2}\left( R^{4}-\frac{4}{3}\Theta^{4}\right)(dX^{+})^{2} -2(R^{2}-\Theta^{2})dX^{+}dX^{-}\nonumber\\
-\mu^{2}\kappa^{2}(R^{4}-\Theta^{4})(dX^{+})^{2}
\end{eqnarray}
in the metric near the Penrose limit at leading order in $ L^{-2} $. 

Before we proceed further, it is indeed worthwhile to explore the above background (\ref{E11}) in the limit, $ \kappa \rightarrow i $ with a proper rescaling of the coordinates namely,
\begin{eqnarray}
R\rightarrow\frac{R}{\sqrt{\varepsilon}}, ~~\Theta \rightarrow \frac{\Theta}{\sqrt{\varepsilon}},~~|\varepsilon | \ll 1
\end{eqnarray}
such that the corresponding metric in the pp wave limit turns out to be,
\begin{eqnarray}
ds^{2}=-4 dX^{+}dX^{-}- \mu ^2 (\mathcal{Z}^{I})^{2}(dX^{+})^2 +\frac{1}{\varepsilon}(d\mathcal{Z}^{I})^{2},~~I=1,..8.
\end{eqnarray}

We now focus on the NS-NS sector of the theory and express the two forms as \footnote{As we shall see shortly, the NS-NS sector associated with the $ \eta $ deformations is a total derivative term and therefore does not contribute to the stringy dynamics in the strict pp wave limit. It contributes only in the next to leading order in the pp wave limit.},
\begin{eqnarray*}
B_{2AdS_5}=\frac{\kappa R^{4}}{L^{2}}\cos \zeta \sin \zeta d\psi_{1}\wedge d\zeta +\mu \kappa R\left(1+\frac{\kappa^{2}R^{2}}{L^{2}} \right)  dX^{+} \wedge dR \nonumber\\
+\frac{\kappa R}{\mu L^{2}}dX^{-}\wedge dR +\mathcal{O}(1/L^{4})
\end{eqnarray*}
\begin{eqnarray}
B_{2S^5}=-\frac{\kappa \Theta^{4}}{L^{2}}\cos \xi \sin \xi d\phi_{1}\wedge d\xi -\mu \kappa \Theta \left( 1-\frac{\Theta^{2}}{3L^{2}}(2+3 \kappa ^2)\right) dX^{+}\wedge d\Theta \nonumber\\+\frac{\kappa \Theta}{\mu L^{2}}dX^{-}\wedge d\Theta
 +\mathcal{O}(1/L^{4}).\label{e35}
\end{eqnarray}

Before we conclude this Section, a few important points are worthwhile to be mentioned. It should be noticed that at leading order in the Penrose limit, the $ \kappa $ deformation does not add anything non trivially to the stringy spectrum except for the rescaling of the mass parameter ($ \mu $) of the theory. On the other hand, the background deformations start contributing non trivially only at subleading order in the large $ L $ expansion. Therefore, in order to find non trivial contributions due to $ \kappa $ deformations one must go beyond the pp wave limit and compute the Hamiltonian at subleading order in $ 1/L $ expansion. However, before we proceed further, it is always instructive to explore the standard light cone quantization \cite{Plefka:2003nb} for strings propagating over $ \kappa $ deformed $ AdS_5 \times S^{5} $.
%%%%%%%%%%%%%%%%%%%%%%%%%%%%%%%%%%%%%%%%%%%%%%%%%%%
\subsubsection{Light cone quantization}
Before we proceed further, it is important re-emphasize the connection between the conserved quantities (associated with the so called global coordinates ($ t , \psi $)) namely the energy ($ E=i \partial_{t} $) and the angular momentum ($ J =-i\partial_{\psi} $) to that with the newly introduced light cone coordinates ($ X^{\pm} $) and their conjugate variables. The variable conjugate to $ X^{+} $ could be formally expressed as,
\begin{eqnarray}
H_{lc}=i \partial_{X^{+}}=\frac{2\hat{\mu}}{\sqrt{1+ \kappa^{2}}} (i\partial_{t}+i \partial_{\psi})\equiv \frac{\hat{\mu}}{\sqrt{1+ \kappa^{2}}} (E-J)\label{ppwave}
\end{eqnarray}
which in turn implies that for $ \eta $ deformed strings the corresponding light-cone Hamiltonian could be obtained after a proper rescaling of the following form,
\begin{eqnarray}
\hat{\mathcal{H}}_{lc}=\sqrt{1+\kappa^{2}}H_{lc}=\hat{\mu}(E-J).
\end{eqnarray}

On the other hand, the other conjugate momenta could be formally expressed as,
\begin{eqnarray}
2P^{+}=-P_{-}=i\partial_{X^{-}}=\frac{2\sqrt{1+\kappa^{2}}}{\hat{\mu} L^{2}}(E+J).
\end{eqnarray}
Clearly, one should notice that in order to define states with non zero momenta $ P^{+} $, one must therefore set the angular momentum to be large namely, $ J\sim L^{2} $ in the corresponding pp wave limit.

The purpose of this Section is to explore the so called Light cone quantization corresponding to bosonic string theory on deformed $ AdS_5 \times S^{5} $ in the presence of $ NS $ fluxes. In our analysis, we focus over the background (\ref{E11}),
\begin{eqnarray}
ds^{2}&=&-4 dX^{+}dX^{-}-\hat{\mu}^2 ((\mathcal{Z}^{i}_{R})^{2} +(\mathcal{Z}^{m}_{\Theta})^{2})(dX^{+})^2+(d\mathcal{Z}^{i}_{R})^{2}+(d\mathcal{Z}^{m}_{\Theta})^{2}\nonumber\\
B_{2}&=&\mu \kappa (\mathcal{Z}^{i}_{R}dX^{+}\wedge d\mathcal{Z}^{i}_{R}- \mathcal{Z}^{m}_{\Theta}dX^{+}\wedge d\mathcal{Z}^{m}_{\Theta})
\end{eqnarray}
obtained by taking a Penrose limit to the original metric which is essentially the geometry seen by the boosted string travelling along a light like trajectory passing through an equator of the deformed $ S^{5} $. Notice that, here the Cartesian coordinates, $ \lbrace\mathcal{Z}^{i,m}_{R,\Theta} \rbrace$ with $ i=1,..,4 $ and, $ m=5,..,8 $ span the eight dimensional space transverse to the null geodesic. As a consequence of this, like in the $ AdS_5 \times S^{5} $ example \cite{Plefka:2003nb}, the original $ SO(8) $ symmetry is broken down to two copies of $ SO(4) $. 

Typically, in a light-cone quantization one starts with the following gauge namely,
\begin{eqnarray}
X^{+}=\tau\label{E17}
\end{eqnarray}
which eventually eliminates the light cone variables $ X^{\pm} $ and projects the theory on the eight dimensional transverse space.

Our next task would be to construct the action,
\begin{eqnarray}
S_{P}=\frac{\sqrt{\hat{\lambda}}}{2 \pi}\int d^{2}\sigma \mathcal{L}_{P}
\end{eqnarray}
corresponding to the bosonic sector of the full superstring theory. Notice that, here $\hat{\lambda} =\frac{L^{4}}{\alpha'^{2}} (1+\kappa^{2})$ is the modified t'Hooft coupling in the theory with $ \eta $ deformations \cite{Hoare:2014pna}. 
The corresponding Polyakov Lagrangian could be formally expressed as\footnote{Here, we have used the following notation namely, $\dot{\mathcal{Z}}= \partial_{\tau}\mathcal{Z} $ and $ \mathcal{Z}'=  \partial_{\sigma}\mathcal{Z} $.},
\begin{eqnarray}
\mathcal{L}_{P}&=&(\dot{\mathcal{Z}^{i}_{R}})^{2}+(\dot{\mathcal{Z}^{m}_{\Theta}})^{2}-(\mathcal{Z}^{'i}_{R})^{2}-(\mathcal{Z}^{'m}_{\Theta})^{2}-\hat{\mu}^{2}((\mathcal{Z}^{i}_{R})^{2}+(\mathcal{Z}^{m}_{\Theta})^{2})-\mu \kappa \mathcal{Z}^{i}_{R}\mathcal{Z}^{'i}_{R}+\mu \kappa \mathcal{Z}^{m}_{\Theta}\mathcal{Z}^{'m}_{\Theta}\nonumber\\
&=& \mathcal{L}_{R}+\mathcal{L}_{\Theta}
\end{eqnarray}
that essentially decouples into two mutually \textit{non} interacting Lagrangians corresponding to two massive scalar fields. Therefore for the purpose of our present analysis, it is in fact sufficient two focus on one of these scalar sectors where the other sector essentially follows quite trivially.

We focus on the Lagrangian of the following form,
\begin{eqnarray}
\mathcal{L}_{P}=\frac{1}{2}(\dot{\mathcal{Z}}^{2}_{a}-\mathcal{Z}^{'2}_{a}-\hat{\mu}^{2}\mathcal{Z}^{2}_{a}+\gamma_{a}\mu \kappa \mathcal{Z}_{a}\mathcal{Z}'_{a})\label{e22}
\end{eqnarray}
where, we have suppressed the indices $ i ,m$ for the moment and the other index $ a $ runs over two variables namely, $ R $ and $ \Theta $. Here,  $ \gamma_{a} =\pm 1$ depending on which of the scalar sectors we solve. However, one should notice that the last term in (\ref{e22}) is a total derivative term which therefore never appears in the equations of motion.

In the following we note down the corresponding equation of motion, 
\begin{eqnarray}
\ddot{\mathcal{Z}}_{a}-\mathcal{Z}''_{a}+\hat{\mu}^{2}\mathcal{Z}_{a}=0
\end{eqnarray}
where, as we mentioned above, the contribution due to NS fluxes clearly vanishes and one might therefore drop out the indices $ a $ as well. The most general solution in the usual oscillator mode expansion could be formally expressed as, 
\begin{eqnarray}
\mathcal{Z}^{I}(\sigma,\tau)&=&\cos(\hat{\mu}\tau)\frac{\mathfrak{x}^{I}_{0}}{\hat{\mu}}+\sin(\hat{\mu}\tau)\frac{\mathfrak{p}^{I}_{0}}{\hat{\mu}}+\sum_{n\neq 0}\mathfrak{C}_{n}^{I}(\sigma)e^{-j\mathfrak{w}_{n}\tau},~I=1,..,8\nonumber\\
\mathfrak{C}_{n}^{I}(\sigma) &=&\frac{j}{\sqrt{2 \mathfrak{w}_{n}}}(\mathfrak{c}_{n}^{I}e^{j \mathfrak{K}_{n}\sigma}+\mathfrak{d}_{n}^{I}e^{-j\mathfrak{K}_{n}\sigma }),~~j=\sqrt{-1}.\label{E22}
\end{eqnarray}

Imposing the closed string boundary condition,
\begin{eqnarray}
\mathcal{Z}^{I}(\tau ,\sigma)=\mathcal{Z}^{I}(\tau ,\sigma + 2\pi \alpha' P^{+})
\end{eqnarray}
the corresponding dispersion relation could be formally expressed as,
\begin{eqnarray}
\mathfrak{w}_{n}&=&\sqrt{\hat{\mu}^{2}+\mathfrak{K}_{n}^{2}}=\sqrt{\hat{\mu}^{2}+\left( \frac{n}{\alpha' P^{+}}\right) ^{2}}\nonumber\\
& = &\hat{\mu}\sqrt{1+ \frac{\lambda^{(\eta)}_{D} n^{2}}{J^{2}}};~~\lambda^{(\eta)}_{D}=\frac{\hat{\lambda}}{(1+ \kappa^{2})^{2}}=\hat{\lambda} \frac{(1-\eta^{2})^{4}}{(1+\eta^{2})^{4}}\nonumber\\
\mathfrak{K}_{n} &=&\frac{n}{\alpha' P^{+}}\simeq \frac{n}{\alpha'}\frac{\hat{\mu} L^{2}}{J\sqrt{1+\kappa^{2}}}=\frac{n\hat{\mu} \sqrt{\hat{\lambda}}}{J(1+\kappa^{2})}.\label{e26}
\end{eqnarray}
%Like in the $ AdS_5 \times S^{5} $ example \cite{Plefka:2003nb}, one might identify, $ \frac{\lambda^{(\eta)}_{D}}{J^{2}}  $ as being that of the perturbative coupling in the dual gauge theory. 

Remember that (\ref{e26}) is the spectrum created by eight transverse bosonic oscillators acting on the ground state since for each value of $ a $ we have four bosonic creation operators.  The above analysis therefore clearly suggests that atleast for the bosonic sector, the stringy spectrum associated with the $ \eta $ deformed plane wave background is exactly identical to that with the corresponding stringy spectrum associated with the pp wave limit of $ AdS_5 \times S^{5} $ superstring sigma model.

The corresponding canonical momentum could be expressed as,
\begin{eqnarray}
\mathcal{P}^{I}=-\mathfrak{x}^{I}_{0}\sin(\hat{\mu}\tau)+\mathfrak{p}_{0}^{I}\cos(\hat{\mu}\tau)+\sum_{n \neq 0}\sqrt{\frac{\mathfrak{w}_{n}}{2}}(\mathfrak{c}^{I}_{n}e^{-j(\mathfrak{w}_{n}\tau -\mathfrak{K}_{n}\sigma)}+\mathfrak{d}^{I}_{n}e^{-j(\mathfrak{w}_{n}\tau +\mathfrak{K}_{n}\sigma)}).\label{E28}
\end{eqnarray}

Finally, we note down the corresponding canonical Hamiltonian density in the light cone gauge,
\begin{eqnarray}
\mathcal{H}^{pp}_{lc}=\mathcal{P}_{I}\dot{\mathcal{Z}}^{I}-\mathcal{L}_{P}
\end{eqnarray}
which yields the total Hamiltonian associated with the pp wave background,
\begin{eqnarray}
\hat{\mathcal{H}}_{lc}=\int d\sigma ((\mathcal{P}^{I})^{2}+(\partial_{\sigma}\mathcal{Z}^{I})^{2}+\hat{\mu}^{2}(\mathcal{Z}^{I})^{2})=\int d\sigma \mathcal{H}^{pp}_{lc}.\label{e30}
\end{eqnarray}

By introducing creation and annihilation operators one could finally express the stringy spectrum in the light-cone gauge as,
\begin{eqnarray}
\hat{E}_{lc}=\hat{\mu}N_{0}+\hat{\mu}(N_{n}+\tilde{N}_{n})\sqrt{1+ \frac{\lambda^{(\eta)}_{D} n^{2}}{J^{2}}}\label{lightcone}
\end{eqnarray}
where, $ N_0 $ is the occupation number corresponding to zero modes and $ N_n $ and $ \tilde{N}_{n} $ are respectively the number of left movers and the right movers corresponding to non zero modes ($ n \neq 0 $) subjected to the level matching constraint, $ N_{n}=\tilde{N}_{n} $. 
%%%%%%%%%%%%%%%%%%%%%%%%%%%%%%%%%%%%%%%%%%%%%%%%%%%%%%
\subsubsection{A note on the RR sector}
The purpose of this Section is to explore the pp wave limit in the presence of non zero RR couplings, $ \mathcal{F}=e^{\Phi}F $ \cite{Arutyunov:2015mqj}. The presence of the $ RR $ fields modifies the scale invariance equations (\ref{G1}) for the metric and the NS two forms by adding an extra piece that is quadratic in the RR couplings\footnote{Here, $ \mathcal{F}_{i}=e^{\Phi}F_{i} $ are the effective RR couplings that appear in the GS superstring action at quadratic order in fermions.} ($ \mathcal{F}_{i} ~(i=1,3,5)$). The scale invariance equations for RR fields, on the other hand, turn out to be quadratic in the derivative that precisely reduces to that of the standard first order form of type IIB equations in the limit, $ I_{a}=U_{a}=0 $.

For the purpose of our analysis, we would first express the RR one and three form couplings \cite{Arutyunov:2015mqj} in the pp wave limit. A careful analysis reveals,
\begin{eqnarray}
\mathcal{F}_{1}&\sim & \mathcal{O}\left( \frac{1}{L^{4}}\right) \nonumber\\
\mathcal{F}_{3}& \sim &\mathcal{O}\left( \frac{1}{L^{2}}\right)
\end{eqnarray}
which therefore do not contribute in the strict pp wave limit.
On the other hand, the five form field strength turns out to be,
\begin{eqnarray}
\mathcal{F}_{5}=4\mu_{c} \varrho^{3} \sqrt{1+\kappa^{2}}\sin\zeta \cos\zeta dX^{+}\wedge d\psi_{2L}\wedge d\psi_{1L}\wedge d\zeta_{L} \wedge dR \nonumber\\
-4\mu_{c} \theta^{3}\sqrt{1+ \kappa^{2}} \sin\xi \cos\xi dX^{+}\wedge d\phi_{2L}\wedge d\phi_{1L}\wedge d\xi_{L} \wedge d\Theta  + \mathcal{O}(1/L^{2})
\end{eqnarray}
where, $ \mu_{c} =\mu L$ is a dimensionless combination and the entities like $ \psi_{2L} $ and $ \phi_{2L} $ are introduced as new dimensionful coordinates in the pp wave limit. At this stage, it is worthwhile to mention that unlike the case for $ AdS_5 \times S^{5} $ superstrings,  we do not have individual informations on dilaton ($ \Phi $) as well as the RR five form ($ F_{5} $) in the case for $ \eta $ deformed superstrings \cite{Arutyunov:2015mqj}. The only information we have is through the combination, $ e^{\Phi}F_{5} $ which has its source in the quadratic fermionic coupling associated to GS superstrings.
%%%%%%%%%%%%%%%%%%%%%%%%%%%%%%%%%%%%%%%%%%%%%%%%%%%%%
\subsubsection{Remarks on type IIB solutions}
The purpose of this Section is to explore (generalized) type IIB equations \cite{Arutyunov:2015mqj} in the pp wave limit discussed so far. The scale invariance associated with the $ \eta $ deformed strings implies the following $ \beta $ functions corresponding to the metric and the NS-NS two form namely,
\begin{eqnarray}
\beta_{ab}^{\mathfrak{G}}&=&R_{ab}-\frac{1}{4}H_{acd}H_{b}^{cd}- T_{ab}(\mathcal{F}_{i})+D_{a}X_{b}+D_{b}X_{a}\nonumber\\
\beta_{ab}^{\mathfrak{B}}&=&\frac{1}{2}D^{k}H_{kab}+K_{ab}(\mathcal{F}_{i})-X^{k}H_{kab}-D_{a}X_{b}+D_{b}X_{a}\nonumber\\
\beta_{ab}^{X}&=&R -\frac{1}{12}H^{2}_{abc}+4 D_{a}X^{a}-4 X^{a}X_{a}
\end{eqnarray}
where, the functions $ T_{ab} $ and $ K_{ab} $ are precisely the contributions appearing from the RR couplings associated with the GS action \cite{Arutyunov:2015mqj}. 

A careful analysis reveals that in the pp wave limit,
\begin{eqnarray}
X = \frac{4 \kappa \hat{\mu}}{\sqrt{1+\kappa^{2}}} dX^{+}+\mathcal{O}(1/L^{2})\label{X}.
\end{eqnarray} 

Using (\ref{X}), and based on our analysis in the previous Sections, it is quite obvious to note down the following,
\begin{eqnarray}
\beta_{ab}^{\mathfrak{G}}&\simeq &R_{ab}- \mathcal{F}_{acdef}\mathcal{F}_{b}\ ^{cdef}+ \mathcal{O}(1/L^{2})\nonumber\\
\beta_{ab}^{\mathfrak{B}}&\sim & \mathcal{O}(1/L^{2})\nonumber\\
\beta_{ab}^{X}& \sim &\mathcal{O}(1/L^{2})
\end{eqnarray}
which reduce to that of the ordinary type IIB equations in the presence of RR five form coupling, $ \mathcal{F}_{5} $.
%%%%%%%%%%%%%%%%%%%%%%%%%%%%%%%%%%%%%%%%
\subsection{Beyond pp wave limit}
The purpose of this Section is to derive the light-cone Hamiltonian corresponding to the bosonic strings propagating over the background (\ref{E11}) and to calculate the $ 1/L^{2} $ corrections to the light-cone Hamiltonian. This will eventually determine the first order shift in the energy spectrum (\ref{e26}) due to curvature corrections near the Penrose limit.

In order to do that, we first rewrite the background metric (\ref{E11}) in the following form namely,
\begin{eqnarray}
ds^{2}=\mathfrak{G}_{+-}dX^{+}dX^{-}+\mathfrak{G}_{++}(dX^{+})^{2}+\mathfrak{G}^{(R)}_{ij}d\mathcal{Z}^{i}_{R}d\mathcal{Z}^{j}_{R}+\mathfrak{G}^{(\Theta)}_{mn}d\mathcal{Z}^{m}_{\Theta}d\mathcal{Z}^{n}_{\Theta}+\mathcal{O}(1/L^{4})
\end{eqnarray}
where, the individual metric components could be formally expressed as,
\begin{eqnarray*}
\mathfrak{G}_{+-}=-4-\frac{2}{L^{2}}(1+\kappa^{2})(\mathcal{Z}^{I})^{2},~~(\mathcal{Z}^{I})^{2}=(\mathcal{Z}^{i}_{R})^{2}+(\mathcal{Z}^{m}_{\Theta})^{2}
\end{eqnarray*}
\begin{eqnarray*}
\mathfrak{G}_{++}=-\hat{\mu}^{2}(\mathcal{Z}^{I})^{2}+\frac{\mu^{2}}{3L^{2}}(\mathcal{Z}^{m}_{\Theta})^{2}(\mathcal{Z}^{n}_{\Theta})^{2}-\frac{\hat{\mu}^{2}\kappa^{2}}{L^{2}}(\mathcal{Z}^{i}_{R})^{2}(\mathcal{Z}^{j}_{R})^{2}\nonumber\\+\frac{\mu^{2}\kappa^{2}}{L^{2}}\left(\frac{4}{3}+\kappa^{2} \right)(\mathcal{Z}^{m}_{\Theta})^{2}(\mathcal{Z}^{n}_{\Theta})^{2} 
\end{eqnarray*}
\begin{eqnarray*}
\mathfrak{G}^{(R)}_{ij}=\delta_{ij}\left(1 -\frac{(1-\kappa^{2})}{L^{2}}\mathcal{Z}^{i}_{R}\mathcal{Z}^{j}_{R}\right) -\frac{(1-\kappa^{2})}{L^{2}}\mathcal{Z}^{i}_{R}\mathcal{Z}^{j}_{R} ~,~~i,j=1,..,4
\end{eqnarray*}
\begin{eqnarray}
\mathfrak{G}^{(\Theta)}_{mn}=\delta_{mn}\left( 1 -\frac{1}{3 L^{2}}((\mathcal{Z}^{k}_{\Theta})^{2}-\mathcal{Z}^{m}_{\Theta}\mathcal{Z}^{n}_{\Theta})-\frac{\kappa^{2}}{L^{2}}\mathcal{Z}^{m}_{\Theta}\mathcal{Z}^{n}_{\Theta}\right)\nonumber\\ +\frac{1}{L^{2}}\left( \frac{1}{3}-\kappa^{2}\right) (\mathcal{Z}^{m}_{\Theta})(\mathcal{Z}^{n}_{\Theta})~,~m,n=5,..,8.\label{E26}
\end{eqnarray}

Our next task would be to express the NS fluxes (\ref{e35}) in an appropriate basis of $ \lbrace X^{\pm}, \mathcal{Z}^{i}_{R},\mathcal{Z}^{m}_{\Theta}\rbrace $. In order to do that, we first note down the mapping between the angular and the Cartesian ($ \mathcal{Z}^{i}_{R} $) variables namely\footnote{Similar arguments follow quite trivially for the other set of variables $ \lbrace \mathcal{Z}^{m}_{\Theta}\rbrace $ where, $ m=5,6,7,8 $.},
\begin{eqnarray}
\mathcal{Z}^{1}_{R}&=&\cos \psi_{1} \sin\zeta ,~\mathcal{Z}^{2}_{R}=\sin \psi_{1} \sin\zeta \nonumber\\
\mathcal{Z}^{3}_{R}&=&\cos \psi_{2} \cos\zeta ,~\mathcal{Z}^{4}_{R}=\sin \psi_{2} \cos\zeta\label{e52}
\end{eqnarray}
where, we have set the radius of the three sphere, $ R=\sqrt{\mathcal{Z}^{i}_{R} \mathcal{Z}^{i}_{R}}=1 $. From (\ref{e52}) it is trivial to find the following mapping between the coordinates,
\begin{eqnarray}
\zeta &=& \tan^{-1}\sqrt{\frac{(\mathcal{Z}^{1}_{R})^{2}+(\mathcal{Z}^{2}_{R})^{2}}{(\mathcal{Z}^{3}_{R})^{2}+(\mathcal{Z}^{4}_{R})^{2}}}=\Pi^{(R)} (\mathcal{Z}^{1}_{R}, \mathcal{Z}^{2}_{R},\mathcal{Z}^{3}_{R}, \mathcal{Z}^{4}_{R})\nonumber\\
\psi_{1}&=&\tan^{-1}\left( \frac{\mathcal{Z}^{2}_{R}}{\mathcal{Z}^{1}_{R}}\right) =\Omega^{(R)} (\mathcal{Z}^{1}_{R}, \mathcal{Z}^{2}_{R}).\label{e53}
\end{eqnarray}

A similar analysis reveals,
\begin{eqnarray}
\xi &=& \tan^{-1}\sqrt{\frac{(\mathcal{Z}^{5}_{\Theta})^{2}+(\mathcal{Z}^{6}_{\Theta})^{2}}{(\mathcal{Z}^{7}_{\Theta})^{2}+(\mathcal{Z}^{8}_{\Theta})^{2}}}=\Pi^{(\Theta)} (\mathcal{Z}^{5}_{\Theta}, \mathcal{Z}^{6}_{\Theta},\mathcal{Z}^{7}_{\Theta}, \mathcal{Z}^{8}_{\Theta})\nonumber\\
\phi_{1}&=&\tan^{-1}\left( \frac{\mathcal{Z}^{6}_{\Theta}}{\mathcal{Z}^{5}_{\Theta}}\right) =\Omega^{(\Theta)} (\mathcal{Z}^{5}_{\Theta}, \mathcal{Z}^{6}_{\Theta}).\label{e54}
\end{eqnarray}

Using (\ref{e53}) and (\ref{e54}), the NS fluxes could be formally expressed as,
\begin{eqnarray*}
B_{2AdS_{5}}=\frac{\kappa}{L^{2}}(\mathcal{Z}^{l}_{R})^{2}(\mathcal{Z}^{j}_{R})^{2}\cos \Pi^{(R)} \sin \Pi^{(R)} \partial_{a}\Omega^{(R)} \partial_{i}\Pi^{(R)} d\mathcal{Z}^{a}_{R}\wedge d\mathcal{Z}^{i}_{R}\nonumber\\ +\mu \kappa \mathcal{Z}^{i}_{R}\left(1+\frac{\kappa^{2}(\mathcal{Z}^{j}_{R})^{2}}{L^{2}} \right)  dX^{+} \wedge d\mathcal{Z}^{i}_{R} 
+\frac{\kappa }{\mu L^{2}}\mathcal{Z}^{i}_{R}dX^{-}\wedge d\mathcal{Z}^{i}_{R} +\mathcal{O}(1/L^{4})
\end{eqnarray*}
\begin{eqnarray}
B_{2S^5}=-\frac{\kappa }{L^{2}}(\mathcal{Z}^{p}_{\Theta})^{2}(\mathcal{Z}^{n}_{\Theta})^{2}\cos \Pi^{(\Theta)} \sin \Pi^{(\Theta)} \partial_{k}\Omega^{(\Theta)} \partial_{m}\Pi^{(\Theta)} d\mathcal{Z}^{k}_{\Theta}\wedge d\mathcal{Z}^{m}_{\Theta}\nonumber\\-\mu \kappa \mathcal{Z}^{m}_{\Theta} \left( 1-\frac{(\mathcal{Z}^{n}_{\Theta})^{2}}{3L^{2}}(2+3 \kappa ^2)\right) dX^{+}\wedge d\mathcal{Z}^{m}_{\Theta} +\frac{\kappa}{\mu L^{2}}\mathcal{Z}^{m}_{\Theta}dX^{-}\wedge d\mathcal{Z}^{m}_{\Theta}
 +\mathcal{O}(1/L^{4})\label{E27}
\end{eqnarray}
where, the indices, $ a=1,2 $,  $ i,j=1,2,3,4 $, $ k=5,6 $ and $ m,n=5,6,7,8 $ are perfectly consistent with our previous notations.

Our next task would be to use (\ref{E26}, \ref{E27}) inorder to calculate the curvature corrections to the world sheet Hamiltonian
in the light-cone gauge (\ref{E17}). 
We start with the Polyakov Lagrangian \cite{Callan:2004uv} of the following form,
\begin{eqnarray}
\mathcal{L}_{P}=\frac{1}{2}(h^{ab}\partial_{a}X^{\mu}\partial_{b}X^{\nu}\mathfrak{G}_{\mu \nu}-\epsilon^{ab}\partial_{a}X^{\mu}\partial_{b}X^{\nu}\mathfrak{B}_{\mu \nu})
\end{eqnarray}
where, the indices ($ a,b $) label the world-sheet coordinates $ \sigma $ and $ \tau $. Here, $ \epsilon^{ab} $ is the antisymmetric tensor with the convention, $ \epsilon^{\tau \sigma}=-\epsilon^{\sigma \tau}=1 $. Moreover, here $ h^{ab}=\sqrt{-\gamma}\gamma^{ab} $ is built out of the world-sheet metric such that, $ \sqrt{-\gamma}=1 $ and there are only two independent components of the metric\footnote{At this stage it is noteworthy to point out that the conformal gauge condition namely choosing the world sheet metric to be flat, $ h^{ab}=(-1,+1) $ is not a consistent choice beyond the pp wave limit. It turns out that in order to maintain the light-cone gauge (\ref{E17}) one must therefore take into account curvature square corrections to the world-sheet metric $ h^{ab} $ \cite{Callan:2004uv}.}.

The canonical momentum is defined as,
\begin{eqnarray}
\mathcal{P}_{\mu}=\frac{\partial \mathcal{L}_{P}}{\partial \dot{X}^{\mu}}=h^{a \tau}\partial_{a}X^{\nu}\mathfrak{G}_{\mu \nu}-\epsilon^{ \tau a}\partial_{a}X^{\beta}\mathfrak{B}_{\mu \beta}\label{E29}
\end{eqnarray}
which could be solved in order to obtain,
\begin{eqnarray}
\dot{X}^{\mu}=\frac{1}{h^{\tau \tau}}\mathfrak{G}^{\mu \nu}\left( \mathcal{P}_{\nu}+\mathfrak{B}_{\nu \beta}X'^{\beta}\right) -\frac{h^{\tau \sigma}}{h^{\tau \tau}}X'^{\mu}.\label{E30}
\end{eqnarray}

Using (\ref{E30}), the canonical Hamiltonian,
\begin{eqnarray}
\mathcal{H}_{lc}=\mathcal{P}_{\mu}\dot{X}^{\mu}-\mathcal{L}_{P}
\end{eqnarray}
could be formally expressed as,
\begin{eqnarray}
\mathcal{H}_{lc}=\frac{1}{2h^{\tau \tau}}\left(\mathcal{P}_{\mu}\mathfrak{G}^{\mu \nu} \mathcal{P}_{\nu}+X'^{\mu}\mathfrak{G}_{\mu \nu}X'^{\nu} +\mathfrak{G}^{\mu \alpha} \mathfrak{B}_{\alpha \lambda}\mathfrak{B}_{\mu \nu}X'^{\lambda}X'^{\nu}\right) \nonumber\\
-\frac{h^{\tau \sigma}}{h^{\tau \tau}}\mathcal{P}_{\mu}X'^{\mu}+\frac{1}{h^{\tau \tau}}\mathfrak{G}^{\mu \alpha}\mathcal{P}_{\alpha}X'^{\nu}\mathfrak{B}_{\mu \nu}.
\end{eqnarray}

Our next task would be to consider Virasoro constraints,
\begin{eqnarray}
T_{\tau \tau}&=&\frac{1}{2}(h^{\tau \tau})^{2}\dot{X}^\mu\dot{X}^{\nu}\mathfrak{G}_{\mu \nu}+h^{\tau \sigma}h^{\tau \tau}\dot{X}^{\mu}X'^{\nu}\mathfrak{G}_{\mu \nu}+\left(1+\frac{1}{2}h^{\tau \tau}h^{\sigma \sigma} \right) X'^{\mu}X'^{\nu}\mathfrak{G}_{\mu \nu}=0\nonumber\\
T_{\tau \sigma}&=&\frac{1}{2} h^{\tau \sigma}\left(h^{\tau \tau}\dot{X}^{\mu}\dot{X}^{\nu}+h^{\sigma \sigma}X'^{\mu}X'^{\nu}\right)\mathfrak{G}_{\mu \nu} +h^{\tau \tau}h^{\sigma \sigma}\dot{X}^{\mu}X'^{\nu}\mathfrak{G}_{\mu \nu}=0.\label{E33}
\end{eqnarray}

Using (\ref{E30}), one could in fact rewrite the above constraints (\ref{E33}) as,
\begin{eqnarray}
\mathcal{P}_{\mu}\mathfrak{G}^{\mu \nu}\mathcal{P}_{\nu}+2\mathcal{P}_{\mu}\mathfrak{G}^{\mu \lambda}\mathfrak{B}_{\lambda \nu}X'^{\nu}+\mathfrak{G}^{\mu \nu}\mathfrak{B}_{\mu \lambda}\mathfrak{B}_{\nu \beta}X'^{\lambda}X'^{\beta}+X'^{\mu}X'^{\nu}\mathfrak{G}_{\mu \nu}&=&0\nonumber\\
\mathcal{P}_{\mu}X'^{\mu}&=&0.\label{E59}
\end{eqnarray}

In order to proceed further, we choose the light cone gauge \cite{Callan:2004uv},
\begin{eqnarray}
X^{+}=\tau ,~~P_{-}=C
\end{eqnarray}
and split the constraints (\ref{E59}) explicitly into light cone coordinates which yields the following set of equations,
\begin{eqnarray}
P_{-}X'^{-}+\mathcal{P}_{I}\mathcal{Z}^{'I}=0,\label{E64}
\end{eqnarray}
\begin{eqnarray}
\mathfrak{G}^{++}P^{2}_{+}+2 \mathfrak{G}^{+-}P_{+}P_{-}+\mathfrak{G}^{--}P^{2}_{-}+\mathfrak{G}^{ij}\mathcal{P}_{i}\mathcal{P}_{j}+\mathfrak{G}^{mn}\mathcal{P}_{m}\mathcal{P}_{n}
+2P_{+}\mathfrak{G}^{++}\mathfrak{B}_{+I}\mathcal{Z}'^{I}\nonumber\\
2P_{+}\mathfrak{G}^{+-}\mathfrak{B}_{-I}\mathcal{Z}'^{I}+2P_{-}\mathfrak{G}^{-+}\mathfrak{B}_{+I}\mathcal{Z}'^{I}+2P_{-}\mathfrak{G}^{--}\mathfrak{B}_{-I}\mathcal{Z}'^{I}
+2\mathcal{P}_{i}\mathfrak{G}^{ij}\mathfrak{B}_{ja}\mathcal{Z}^{'a}_{R}+2\mathcal{P}_{m}\mathfrak{G}^{mn}\mathfrak{B}_{nk}\mathcal{Z}^{'k}_{\Theta}\nonumber\\
+(\mathfrak{G}^{++}\mathfrak{B}_{+I}\mathfrak{B}_{+J}+2\mathfrak{G}^{+-}\mathfrak{B}_{+I}\mathfrak{B}_{-J})\mathcal{Z}'^{I}\mathcal{Z}'^{J}+\mathfrak{G}^{IJ}\mathfrak{B}_{I-}\mathfrak{B}_{J-}(X'^{-})^{2}+2\mathfrak{G}^{ij}\mathfrak{B}_{ia}\mathfrak{B}_{j-}X'^{-}\mathcal{Z}'^{a}_{R}\nonumber\\
+\mathfrak{G}^{ij}\mathfrak{B}_{ia}\mathfrak{B}_{jb}\mathcal{Z}'^{a}_{R}\mathcal{Z}'^{b}_{R}+2\mathfrak{G}^{mn}\mathfrak{B}_{mk}\mathfrak{B}_{n-}X'^{-}\mathcal{Z}'^{k}_{\Theta}
+\mathfrak{G}^{mn}\mathfrak{B}_{mk}\mathfrak{B}_{np}\mathcal{Z}'^{k}_{\Theta}\mathcal{Z}'^{p}_{\Theta}\nonumber\\
+\mathcal{Z}'^{i}_{R}\mathcal{Z}'^{j}_{R}\mathfrak{G}_{ij}+\mathcal{Z}'^{m}_{\Theta}\mathcal{Z}'^{n}_{\Theta}\mathfrak{G}_{mn}=0\label{E65}
\end{eqnarray}
where the indices, $ I,J=1,..,8 $ run over all the eight spatial indices. All the remaining indices are already defined below (\ref{E27}). At this stage, it is noteworthy to mention that the terms proportional to $ \mathcal{B}_{\pm I} $ are not important as far as the dynamics is concerned. This is due to the fact that these correspond to the near pp wave expansion of the total derivative term in the $ B $ field (\ref{e19}) expressed in the light cone coordinates. Therefore, for the purpose of our rest of the analysis we simply drop these terms.

It is now quite trivial to solve $ X'^{-} $ from (\ref{E64}) and substitute it back into (\ref{E65}) which finally yields the light-cone Hamiltonian in the so called transverse coordinates. However, one should notice that for $ \mathfrak{G}_{--}=0 $ the above equation becomes linear \cite{Callan:2004uv} in the light-cone momentum ($ P_+ $) which finally yields\footnote{One should notice that for $ \eta $ deformation, the background metric, $ \mathfrak{G}_{--} $ is identically zero at order $ 1/L^{2} $. This is similar to that of the $ AdS_5 \times S^{5} $ background \cite{Callan:2004uv}.},
\begin{eqnarray}
\mathcal{H}_{lc}=-\frac{(\mathfrak{G}^{--}C^{2}+\mathcal{K}(\mathcal{Z}^{I},\mathcal{P}^{I}))}{2\mathfrak{G}^{+-}C}\label{E66}
\end{eqnarray}
which clearly maps to that of the light-cone Hamiltonian of the $ AdS_5 \times S^{5} $ strings \cite{Asano:2015qwa} in the limit of the vanishing deformation, $ \kappa \rightarrow 0 $. The function, $ \mathcal{K}(\mathcal{Z^{I}},\mathcal{P}^{I}) $ on the other hand, could be easily read off from (\ref{E65}) which essentially contains terms both linear as well as quadratic in the $ B $ field. A careful analysis reveals that terms quadratic in the $ B $ field always contribute at higher orders in $ 1/L^{2} $. Therefore these terms are not important at next to leading order in the near Penrose limit. The reason for this rests on the fact that $ B $ fields start contributing only beyond the Penrose limit. In the Penrose limit, as we have seen previously, the $ B $ field contributions to the Polyakov action appear as a total derivative  and hence they are not important. A careful analysis reveals\footnote{The exact expression for the function $ \Lambda (\mathcal{Z}^{I},\mathcal{P}^{I}) $ has been provided in the Appendix.},
\begin{eqnarray}
\mathcal{K}(\mathcal{Z}^{I},\mathcal{P}^{I})=(\mathcal{P}^{I})^{2}++(\mathcal{Z}'^{I})^{2}+\frac{1}{L^{2}}\Lambda (\mathcal{Z}^{I},\mathcal{P}^{I})+\mathcal{O}(1/L^{4})
\end{eqnarray}
which could be further used in order to re-express (\ref{E66}) as,
\begin{eqnarray}
\mathcal{H}_{lc}=\mathcal{H}^{pp}_{lc}+\frac{2}{L^{2}}\Delta \mathcal{H}+\mathcal{O}(1/L^{4})\label{e68}
\end{eqnarray}
where, the leading order curvature correction near the pp wave limit could be formally expressed as,
\begin{eqnarray}
\Delta \mathcal{H}=\frac{(1+\kappa^{2})}{4}((\mathcal{P}^{I})^{2}+(\mathcal{Z}'^{I})^{2})(\mathcal{Z}^{I})^{2}+\frac{\Lambda}{2}-\frac{\mu^{2}}{4}(1+\kappa^{2})^{2}(\mathcal{Z}^{I})^{4}-\frac{1}{2}\mathcal{G}_{PP}\nonumber\\
\mathcal{G}_{PP}=\frac{\mu^{2}}{3}(\mathcal{Z}^{m}_{\Theta})^{2}(\mathcal{Z}^{n}_{\Theta})^{2}-\hat{\mu}^{2}\kappa^{2}(\mathcal{Z}^{i}_{R})^{2}(\mathcal{Z}^{j}_{R})^{2}+\mu^{2}\kappa^{2}\left(\frac{4}{3}+\kappa^{2} \right)(\mathcal{Z}^{m}_{\Theta})^{2}(\mathcal{Z}^{n}_{\Theta})^{2} 
\end{eqnarray}
subjected to the fact that we have set the constant, $ C= 4$. Here, $ \mathcal{H}^{pp}_{lc} $ corresponds to the usual light-cone Hamiltonian (\ref{e30}) associated to the sigma model in the pp wave limit. Once expressed in terms of the creation and annihilation operators, the leading curvature square ($ 1/L^{2} $) correction to the above bosonic Hamiltonian (\ref{e68}) would correspond to $ 1/J $ corrections to the string pp wave spectrum. However, this is not the end of the story since the fermionic counterpart for $ \eta $ deformed superstrings are yet to be explored.  Finally, one should also notice that like in the case for the $ AdS_5 \times S^{5} $ superstrings, adding corrections beyond $ 1/L^{2} $ regime might lead us towards a more delicate situation where one might encounter the problem of dealing with the square root of the Hamiltonian due to non vanishing metric component $ \mathfrak{G}_{--} $. 
%%%%%%%%%%%%%%%%%%%%%%%%%%%%%%%%%%%%%%%%%%%%%%%%%%%%%%%%%
\section{A note on the uniform gauge}
%%%%%%%%%%%%%%%%%%%%%%%%%%%%%%%%%%%%%%%%%%
\subsection{The Hamiltonian formulation}
%%%%%%%%%%%%%%%%%%%%%%%%%%%%%%%%%%%%%%
The purpose of this part of the analysis is to explore the Hamiltonian formulation for the $ \eta $ deformed strings in the so called uniform gauge \cite{Arutyunov:2004yx} and to check the corresponding BMN limit. To start with, we consider the following change of coordinates,
\begin{eqnarray}
\varrho =\sinh \rho
\end{eqnarray}
which yields the deformed $ AdS_5 $ part of the metric as,
\begin{eqnarray}
ds^{2}_{ (AdS_{5})_{\kappa}}= -\frac{\cosh^{2}\rho}{(1-\kappa^2 \sinh^{2}\rho)}dt^2+\frac{d\rho^{2}}{(1- \kappa^{2}\sinh^{2}\rho)}+\frac{\sinh^{2}\rho d\zeta^{2}}{1+\kappa^{2}\sinh^{4}\rho\sin^{2}\zeta}\nonumber\\
+\frac{\sinh^{2}\rho\cos^{2}\zeta}{1+\kappa^{2}\sinh^{4}\rho\sin^{2}\zeta} d\psi^{2}_{1}+\sinh^{2}\rho\sin^{2}\zeta d\psi^{2}_{2}.
\end{eqnarray}

Next, we define the following reparametrization \cite{Arutyunov:2004yx},
\begin{eqnarray}
\cosh \rho = \frac{1+\frac{z^{2}}{4}}{1-\frac{z^{2}}{4}}=\mathcal{Q}(z)
\end{eqnarray}
which yields the following metric,
\begin{eqnarray}
ds^{2}_{ (AdS_{5})_{\kappa}}&\simeq & -\frac{\mathcal{Q}^{2}(z)}{\mathcal{D}(z)}dt^2+\frac{1}{\left(1-\frac{z^{2}}{4}\right) ^{2}}\left( \delta_{ij}+\frac{\kappa^{2}\mathcal{N}^{2}(z)}{z^{2}\mathcal{D}(z)}z_{i}z_{j}\right) dz^{i}dz^{j}+\mathcal{O}(\kappa^{2}\rho^{4})\nonumber\\
& \simeq & -\mathcal{G}_{tt} dt^{2}+\mathcal{G}_{ij}dz^{i}dz^{j}+\mathcal{O}(\kappa^{2}\rho^{4});~i,j=1,..,4\nonumber\\
\mathcal{D}(z)&=&1-\kappa^2 \mathcal{N}^{2}(z)=1-\frac{\kappa^{2}z^{2}}{(1-z^{2}/4)^{2}}\label{E75}
\end{eqnarray}
subjected to the fact that we have boosted the string along the light like geodesic namely, $t= \psi $, $ \rho \sim \theta \sim 0 $ as discussed earlier. This is exactly the pp wave limit considered earlier. However, here we are using a different set of coordinates to parametrize the background solutions. Notice that, with this choice of parametrization, the pp wave limit corresponds to setting, $ z \sim 0$.
Finally, with the choice,
\begin{eqnarray}
\cos \theta = \frac{1-\frac{y^{2}}{4}}{1+\frac{y^{2}}{4}}=\mathcal{R}(y)
\end{eqnarray}
the metric corresponding to the deformed $ S^{5} $ turns out to be,
\begin{eqnarray}
ds^{2}_{ (S^{5})_{\kappa}}&\simeq & \frac{\mathcal{R}^{2}(y)}{\hat{\mathcal{D}}(y)}d\psi^2+\frac{1}{\left(1+\frac{y^{2}}{4}\right) ^{2}}\left( \delta_{mn}-\frac{\kappa^{2}\hat{\mathcal{N}}^{2}(y)}{y^{2}\hat{\mathcal{D}}(y)}y_{m}y_{n}\right) dy^{m}dy^{n}+\mathcal{O}(\kappa^{2}\theta^{4})\nonumber\\
& \simeq & \mathcal{G}_{\psi \psi} d\psi^{2}+\mathcal{G}_{mn}dy^{m}dy^{n}+\mathcal{O}(\kappa^{2}\theta^{4});~m,n=6,..,9\nonumber\\
\hat{\mathcal{D}}(y)&=&1+\kappa^2 \hat{\mathcal{N}}^{2}(y)=1+\frac{\kappa^{2}y^{2}}{(1+y^{2}/4)^{2}}.\label{E77}
\end{eqnarray}
Notice that the NS-NS two forms do not contribute in this limit which is consistent with our previous observation.

The corresponding Polyakov Lagrangian turns out to be,
\begin{eqnarray}
\mathcal{L}_{P}=\frac{\sqrt{\hat{\lambda}}}{2}\gamma^{\alpha \beta}\left(\mathcal{G}_{tt}\partial_{\alpha}t \partial_{\beta}t -\mathcal{G}_{ij}\partial_{\alpha}z^{i}\partial_{\beta}z^{j}-\mathcal{G}_{\psi \psi}\partial_{\alpha}\psi \partial_{\beta}\psi -\mathcal{G}_{mn}\partial_{\alpha}y^{m}\partial_{\beta}y^{n}\right)\label{E78} 
\end{eqnarray}
where, $ \gamma^{\alpha \beta}=\sqrt{-h}h^{\alpha \beta} $ is the world sheet metric in the Minkowski signature and $ \sqrt{\hat{\lambda}}(=\sqrt{\lambda}\sqrt{1+\kappa^{2}}) $ is the $ \kappa $ corrected t'Hooft coupling \cite{Hoare:2014pna} as mentioned previously.

Following the original arguments \cite{Arutyunov:2004yx}, our next step would be to recast the above Lagrangian in terms of canonical momenta. In order to do that, we first note down the canonical momenta as,
\begin{eqnarray}
p_{t}&=&\sqrt{\hat{\lambda}}\mathcal{G}_{tt}(\gamma^{\tau \tau}\dot{t}+\gamma^{\sigma\tau}t')\nonumber\\
p_{z_i}&\equiv &p_{i}=-\sqrt{\hat{\lambda}}\mathcal{G}_{ij}(\gamma^{\tau \tau}\dot{z}^{j}+\gamma^{\sigma\tau}z'^{j})\nonumber\\
p_{\psi}&=&-\sqrt{\hat{\lambda}}\mathcal{G}_{\psi \psi}(\gamma^{\tau \tau}\dot{\psi}+\gamma^{\sigma\tau}\psi')\nonumber\\
p_{z_m}&\equiv &p_{m}=-\sqrt{\hat{\lambda}}\mathcal{G}_{mn}(\gamma^{\tau \tau}\dot{y}^{n}+\gamma^{\sigma\tau}y'^{n}).\label{E79}
\end{eqnarray}

Using (\ref{E79}), one could rewrite the Polyakov Lagrangian (\ref{E78}) as,
\begin{eqnarray}
\mathcal{L}_{P}=p_{t}\dot{t}+p_{i}\dot{z}^{i}+p_{m}\dot{y}^{m}+p_{\psi}\dot{\psi}-\frac{1}{2\sqrt{\hat{\lambda}}\gamma^{\tau \tau}}\left(\frac{p^{2}_{t}}{\mathcal{G}_{tt}} -\mathcal{G}^{ij}p_{i}p_{j}-\mathcal{G}^{mn}p_{m}p_{n}-\frac{p^{2}_{\psi}}{\mathcal{G}_{\psi \psi}}\right)\nonumber\\
-\frac{\sqrt{\hat{\lambda}}}{2\gamma^{\tau \tau}}\left(\mathcal{G}_{tt}t'^{2} -\mathcal{G}_{ij}z'^{i}z'^{j}-\mathcal{G}_{mn}y'^{m}y'^{n}-\mathcal{G}_{\psi \psi}{\psi}^{'2}\right) +\frac{\gamma^{\tau \sigma}}{\gamma^{\tau \tau}}\left( p_{t}t' +p_{\psi}\psi' +p_i z'^{i}+p_{m}y'^{m}\right)\label{E80}
\end{eqnarray}
subjected to the fact, $ det \gamma^{-1}=-1$ together with the inverse metrics as,
\begin{eqnarray}
\mathcal{G}^{ij}&=&\left(1-\frac{z^{2}}{4} \right)^{2}\left( \delta^{ij}-\frac{\kappa^{2}\mathcal{N}^{2}(z)}{z^{2}\mathcal{D}(z)}z^{i}z^{j}\right)\nonumber\\
\mathcal{G}^{mn}&=&\left(1+\frac{y^{2}}{4}\right) ^{2}\left( \delta^{mn}+\frac{\kappa^{2}\hat{\mathcal{N}}^{2}(y)}{y^{2}\hat{\mathcal{D}}(y)}y^{m}y^{n}\right).
\end{eqnarray}

Our next task would be to fix the uniform gauge by imposing the following conditions,
\begin{eqnarray}
t= \tau ,~~p_{\psi}=J \label{E82}
\end{eqnarray}
which eventually maps the space time energy density of strings to that with the Hamiltonian density associated to the physical degrees of freedom of the world-sheet theory,
\begin{eqnarray}
p_{t}=\mathcal{H}.
\end{eqnarray}
Our goal would be to compute $ \mathcal{H} $ in terms of physical degrees of freedom of the world-sheet. In order to do that, we first remove the non-physical d.o.f. of the world-sheet theory by imposing the above gauge fixing conditions\footnote{Upon imposing the gauge fixing (\ref{E82}), the equation of motion corresponding to $ \gamma^{\tau \sigma} $ becomes a constraint equation which further eliminates the non-physical d.o.f. in the system.} (\ref{E82}),
\begin{eqnarray}
\psi' = - \frac{1}{J}(p_i z'^{i}+p_{m}y'^{m})\label{E84}
\end{eqnarray}
which could be integrated further in order to obtain,
\begin{eqnarray}
\psi (2 \pi)-\psi (0)=2 \pi \mathfrak{m}=\frac{1}{J} \int_{0}^{2 \pi}(p_i z'^{i}+p_{m}y'^{m})d \sigma
\end{eqnarray}
where, $ \mathfrak{m} $ is some integer known as the winding number.

Varying the Lagrangian (\ref{E80}) w.r.t. $ \gamma^{\tau \tau} $ yields the square of the Hamiltonian corresponding to $ \eta $ deformed strings,
\begin{eqnarray}
\mathcal{H}^{2}=\frac{\mathcal{G}_{tt}}{\mathcal{G}_{\psi \psi}}J^{2}+\frac{\hat{\lambda}}{J^{2}}\mathcal{G}_{tt}\mathcal{G}_{\psi \psi}\left( \sum_{I=1,I \neq 5}^{9}p_I z'^{I}\right) ^{2}+\frac{\hat{\lambda} \mathcal{G}_{tt}(z'^{i})^{2}}{\left(1-\frac{z^{2}}{4} \right)^{2}}+\mathcal{G}_{tt}\left(1-\frac{z^{2}}{4} \right)^{2}p^{2}_{i}\nonumber\\
+\frac{\mathcal{G}_{tt}\kappa^{2}\mathcal{N}^{2}(z)}{z^{2}\mathcal{D}(\kappa^{2},z)}\left(\frac{\hat{\lambda} (z'^{i}z_i)^{2}}{\left(1-\frac{z^{2}}{4} \right)^{2}} -\left(1-\frac{z^{2}}{4} \right)^{2}(z_i p^{i})^{2}\right) +\frac{\hat{\lambda} \mathcal{G}_{tt}(y'^{m})^{2}}{\left(1+\frac{y^{2}}{4} \right)^{2}}+\mathcal{G}_{tt}\left(1+\frac{y^{2}}{4} \right)^{2}p^{2}_{m}\nonumber\\
-\frac{\mathcal{G}_{tt}\kappa^{2}\hat{\mathcal{N}}^{2}(y)}{y^{2}\hat{\mathcal{D}}(\kappa^{2},y)}\left(\frac{\hat{\lambda}(y'^{m}y_{m})^{2}}{\left(1+\frac{y^{2}}{4} \right)^{2}} -\left(1+\frac{y^{2}}{4} \right)^{2}(y_m p^{m})^{2}\right).\label{E85}
\end{eqnarray}
Clearly, the Hamiltonian is a function of eight (transverse) physical variables ($ z^{i},y^{m} $) once we substitute the constraint (\ref{E84}) in order to eliminate the unphysical degrees of freedom. One should notice that, like in the case for $ AdS_5 \times S^{5} $ superstrings, the background (\ref{E75})-(\ref{E77}) is invariant under $ SO(4)\times SO(4) $ of the full $ R $- symmetry group of the undeformed $ AdS_5 \times S^{5} $  superstring theory. This we have witnessed earlier in the analysis with a different choice of coordinates. As a consequence of this, the resulting Hamiltonian (\ref{E85}) is also invariant under $ SO(4)\times SO(4) $.
%%%%%%%%%%%%%%%%%%%%%%%%%%%%%%%%%%%%%%%%%%%%%%%%%%
\subsection{The BMN limit}
We now focus closely on the BMN limit \cite{Berenstein:2002jq} of the theory. As we noted earlier, the BMN limit is defined through the limit, $ J \rightarrow \infty $ while keeping the ratio, $\tilde{\lambda}=\frac{\lambda}{J^{2}} $ finite and less than unity. In order to make connection with the pp wave Hamiltonian (\ref{e30}), we rescale our coordinates as \cite{Arutyunov:2004yx},
\begin{eqnarray}
z_{i} \rightarrow \frac{z_{i}}{\sqrt{J}},~ y_{m}\rightarrow \frac{y_{m}}{\sqrt{J}},~ p_{i} \rightarrow \sqrt{J}p_{i},~p_{m} \rightarrow \sqrt{J}p_{m}.
\end{eqnarray}

In order to obtain the pp wave Hamiltonian, we expand (\ref{E85}) for the case of short strings with ($ \mathfrak{m}=0 $) zero winding number around the circle parametrized by the angular coordinate, $ \psi $ and keeping the effective coupling, $ \tilde{\lambda} $ finite \cite{Arutyunov:2004yx}. This finally yields,
\begin{eqnarray}
\mathcal{H}=E=J+\mathcal{H}^{pp}+\mathcal{O}(1/J)
\end{eqnarray}
where, the entity,
\begin{eqnarray}
\mathcal{H}^{pp}=\frac{1}{2}\left( p^{2}_{i}+p^{2}_{m}+\frac{\hat{\lambda}}{J^{2}}((z'^{i})^{2}+(y'^{m})^{2})+(1+\kappa^{2})(z^{2}+y^{2})\right) 
\end{eqnarray}
is the corresponding pp wave Hamiltonian density for short strings that reduces to that of \cite{Arutyunov:2004yx} in the limit of the vanishing ($ \kappa \rightarrow 0 $) deformations. The terms $ \mathcal{O}(1/J) $ correspond to corrections appearing beyond the standard pp wave limit that we had discussed previously.

%%%%%%%%%%%%%%%%%%%%%%%%%%%%%%%%%%%%%%%%%%%%%%%%%%
\section{Remarks on HT background}
%%%%%%%%%%%%%%%%%%%%%%%%%%%%%%%%%%%%%%%%%%
\subsection{The pp wave limit}
As mentioned previously, the ABF \cite{Arutyunov:2013ega} background (\ref{e19}) is special in the sense that it is related to an exact type IIB supergravity (HT) background \cite{Hoare:2015wia} via T duality transformation that involves the metric, an imaginary five form and the dilaton. We first note down the metric which could be formally expressed as \cite{Hoare:2015wia},
\begin{eqnarray}
\frac{d\hat{s}^{2}}{L^{2}}=-\frac{1 -\kappa^{2}\varrho^{2}}{1+\varrho^{2}}d \hat{t}^{2}+\frac{d\varrho^{2}}{(1+\varrho^{2})(1- \kappa^{2}\varrho^{2})}+\frac{d \hat{\psi}_{1}^{2}}{\varrho^{2}\cos^{2}\zeta}+(\varrho d\zeta + \kappa \varrho \tan\zeta d\hat{\psi}_{1})^{2}+\frac{d\hat{\phi}^{2}_{2}}{r^{2}\sin^{2}\xi}\nonumber\\
+\frac{d \hat{\psi}_{2}^{2}}{\varrho^{2}\sin^{2}\zeta}+\frac{1+\kappa^{2}r^{2}}{(1-r^{2})}d\hat{\psi}^{2}+\frac{dr^{2}}{(1-r^{2})(1+\kappa^{2}r^{2})}+\frac{d\hat{\phi}^{2}_{1}}{r^{2}\cos^{2}\xi}+(r d \xi -\kappa r\tan \xi d\hat{\phi}_{1})^{2}.
\end{eqnarray}

In order to take the Penrose limit, we consider the following expansion,
\begin{eqnarray}
\hat{t}=\mu  X^{+}+\frac{X^{-}}{\mu L^{2}},~\hat{\psi}=\mu  X^{+}-\frac{X^{-}}{\mu L^{2}},~\hat{\phi}_{1}=\frac{\hat{\Phi}_{1}}{\mu L^{2}},~\hat{\phi}_{2}=\frac{\hat{\Phi}_{2}}{\mu L^{2}}, \nonumber\\
\varrho =\frac{\hat{R}}{L},~r=\frac{\hat{r}}{L},~\zeta =\hat{\zeta} ,~\xi = \hat{\xi},~\hat{\psi}_{1}=\frac{\hat{\Psi}_{1}}{\mu L^{2}},~\hat{\psi}_{2}=\frac{\hat{\Psi}_{2}}{\mu L^{2}}
\end{eqnarray}
which yields the pp wave metric of the following form,
\begin{eqnarray}
d\hat{s}^{2}=-4 dX^{+}dX^{-}+\mu^{2}(1+\kappa^{2})(\hat{R}^{2}+\hat{r}^{2})(dX^{+})^{2}+d\hat{R}^{2}+\hat{R}^{2}d\hat{\zeta}^{2}+d\hat{r}^{2}+\hat{r}^{2}d\hat{\xi}^{2}\nonumber\\
+\frac{d \hat{\Psi}^{2}_{1}}{\mu^{2}\hat{R}^{2}\cos^{2}\hat{\zeta}} +\frac{d \hat{\Psi}^{2}_{2}}{\mu^{2}\hat{R}^{2}\sin^{2}\hat{\zeta}}+\frac{d \hat{\Phi}^{2}_{1}}{\mu^{2}\hat{r}^{2}\cos^{2}\hat{\xi}} +\frac{d \hat{\Phi}^{2}_{2}}{\mu^{2}\hat{r}^{2}\sin^{2}\hat{\xi}}+\mathcal{O}(1/L^{2})\nonumber\\
=-4 dX^{+}dX^{-}+\hat{\mu}^{2}(\hat{\mathcal{Z}_{I}})^{2}(dX^{+})^{2}+(d\mathcal{Z}^{i}_{(\hat{R})})^{2}+(d\mathcal{Z}^{m}_{(\hat{r})})^{2}
+\frac{d \hat{\Psi}^{2}_{1}}{\mu^{2}(\mathcal{Z}^{2}_{(\hat{R})})^{2}}\nonumber\\ +\frac{d \hat{\Psi}^{2}_{2}}{\mu^{2}(\mathcal{Z}^{1}_{(\hat{R})})^{2}}+\frac{d \hat{\Phi}^{2}_{1}}{\mu^{2}(\mathcal{Z}^{4}_{(\hat{r})})^{2}} +\frac{d \hat{\Phi}^{2}_{2}}{\mu^{2}(\mathcal{Z}^{3}_{(\hat{r})})^{2}}+\mathcal{O}(1/L^{2})\label{e54}
\end{eqnarray}
where, $ (\hat{\mathcal{Z}^{I}})^{2} =(\mathcal{Z}^{i}_{(\hat{R})})^{2}+(\mathcal{Z}^{m}_{(\hat{r})})^{2},~(I=1,..4,~i=1,2,~m=3,4)$ is the four dimensional vector that spans the four dimensional space transverse to the remaining six directions .

Next, we note down the dilaton, 
\begin{eqnarray}
\Phi = \Phi_{0}-\log\left( \varrho^{2}r^{2}\sin 2\zeta \sin 2\xi\right) +\mathcal{O}(1/L^{2})\label{e55}
\end{eqnarray}
that eventually turns out to be a regular function in the strict pp wave limit.

Finally, we note down the RR five form coupling of the theory,
\begin{eqnarray}
\mathcal{F}_{5}=\frac{4i r}{\mu_{c} \varrho^{2}\sin \zeta\cos \zeta}dX^{+}\wedge d\hat{\Psi}_{2}\wedge d\hat{\Psi}_{1}\wedge d\hat{\xi}_{L}\wedge d \hat{r}\nonumber\\
-\frac{4i \varrho}{\mu_{c} r^{2}\sin \xi \cos \xi}dX^{+}\wedge d\hat{\Phi}_{2}\wedge d\hat{\Phi}_{1}\wedge d\hat{\zeta}_{L}\wedge d \hat{R}+\mathcal{O}(1/L^{2})
\end{eqnarray}
where, $ \hat{\xi}_{L} $ and $ \hat{\zeta}_{L} $ are the new dimensionful coordinates in the pp wave limit .

%%%%%%%%%%%%%%%%%%%%%%%%%%%%%%%%%%%%%%%%%%
\subsection{Closed string spectrum}
With the above set up in hand, our next task would be to explore the sigma model in the presence of the background metric (\ref{e54}) and the dilaton (\ref{e55}). However, the dilaton term does not contribute once we fix the Weyl invariance by imposing the conformal gauge conditions. The resulting Polyakov Lagrangian turns out to be,
\begin{eqnarray}
\hat{\mathcal{L}}_{P}=(\dot{\hat{\mathcal{Z}_{I}}})^{2}-(\hat{\mathcal{Z}'_{I}})^{2}+\frac{\dot{\hat{\Psi}}_{1}^{2}-\hat{\Psi}_{1}^{'2}}{\mu^{2}(\mathcal{Z}^{2}_{(\hat{R})})^{2}}+\frac{\dot{\hat{\Phi}}_{1}^{2}-\hat{\Phi}_{1}^{'2}}{\mu^{2}(\mathcal{Z}^{4}_{(\hat{r})})^{2}}\nonumber\\
+\frac{\dot{\hat{\Psi}}_{2}^{2}-\hat{\Psi}_{2}^{'2}}{\mu^{2}(\mathcal{Z}^{1}_{(\hat{R})})^{2}}+\frac{\dot{\hat{\Phi}}_{2}^{2}-\hat{\Phi}_{2}^{'2}}{\mu^{2}(\mathcal{Z}^{3}_{(\hat{r})})^{2}}-\hat{\mu}^{2}(\hat{\mathcal{Z}_{I}})^{2} .\label{e57}
\end{eqnarray}

The resulting sets of equations of motion turn out to be \footnote{Here, $ \square = -\partial_{\tau}^{2}+\partial_{\sigma}^{2} $ and the index $ \alpha(=\tau, \sigma) $ runs over the world sheet coordinates such that, $ \eta^{\alpha \beta}=diag(-1,1) $.} \footnote{Here, $ \mathcal{X}^{a} $s stand for the remaining four coordinates transverse to $\hat{ \mathcal{Z}}^{I} $. },
\begin{eqnarray}
(\square -\hat{\mu}^{2})\mathcal{Z}_{I}\delta_{I,n}-\frac{(\dot{\mathcal{X}}^{a})^{2}-(\mathcal{X}'^{a})^{2}}{\mu^{2}(\mathcal{Z}_{n})^{3}}&=&0\nonumber\\
\square \mathcal{X}^{a}-2\partial_{\alpha}\mathcal{X}^{a}\partial^{\alpha}\log \mu \mathcal{Z}_{n} &=&0,
\end{eqnarray}
subjected to the following boundary conditions,
\begin{eqnarray}
\partial_{\sigma}\mathcal{Z}_{n}\delta \mathcal{Z}^{n}|_{\sigma =0}^{\sigma =2\pi}=0,~\frac{1}{\mu^{2}\mathcal{Z}^{2}_{n}}\partial_{\sigma}\mathcal{X}_{a}\delta \mathcal{X}^{a}|_{\sigma =0}^{\sigma =2\pi}=0.
\end{eqnarray}
Here, $ n =1,2,3,4 $ and $ a $ runs over the remaining four indices.
Unlike the case for the $ \eta $ model, the pp wave dynamics associated to HT background turns out to be quite non trivial due to the presence of various self interaction terms in the corresponding Lagrangian (\ref{e57}). In other words, the classical stringy dynamics over HT backgrounds turns out to be highly interacting even in the pp wave limit which makes the theory difficult to solve. However, the dynamics simplifies a lot in the limit, $ |\mu L |\gg 1 $ and one \textit{approximately} recovers the plane wave dispersion relation of the form,
\begin{eqnarray}
\mathfrak{w}^{2}_{n}&\simeq & \frac{n^{2}\lambda}{J^{2}}+\hat{\mu}^{2}\nonumber\\
\mathfrak{w}^{2}_{a}&\simeq & \mathfrak{K}^{2}_{a}.
\end{eqnarray}
 %%%%%%%%%%%%%%%%%%%%%%%%%%%%%%%%%%%%%%%%%%%%%%%%%%%%%%%%%%%%%
\section{Summary and final remarks}
We now conclude our analysis with some final remarks on the implications of the results obtained in this paper. The purpose of the present analysis was to explore the Penrose limit associated with the $ \eta $ deformed string sigma model \cite{Arutyunov:2013ega} and compute the corresponding stringy spectrum for the bosonic sector. Our analysis reveals that as far as the bosonic sector is concerned, except for a trivial rescaling of the mass parameter of the theory, the string sigma model corresponding to $ \eta $ deformations behaves quite analogously to that with the corresponding sigma model associated to $ AdS_5 \times S^{5} $ superstrings in the pp wave limit. It turns out that all the ambiguities associated with the type IIB nature of the background solutions are smoothly washed away and the effective dilaton beta function is zero in the pp wave limit. In order to clarify our results in a better way, we further compute the pp wave Hamiltonian for $ \eta $ deformed strings using the uniform gauge conditions and obtain identical results while probing the sigma model associated with short strings. Finally, we also explore the pp wave limit associated with HT background and explore the corresponding closed string spectrum in that limit. \\ \\
%%%%%%%%%%%%%%%%%%%%%%%%%%%%%%%%%%%%%%%%%%%%%%%%%%%%%%%%%%%%%%%%%%%%%%
{\bf {Acknowledgements :}}
It is a great pleasure to thank Gleb Arutyunov, Oleg Lunin, Rafael Hernandez,  Arkady Tseytlin and Daniel Thompson for their valuable comments on the manuscript. The author would also like convey his sincere thanks to Carlos Nunez and  Salomón Zacarias for various illuminating discussions.
This work was supported through the Newton-Bhahba Fund. The author would like to acknowledge the Royal Society UK and the Science and Engineering Research Board India (SERB) for financial assistance.\\ \\
%%%%%%%%%%%%%%%%%%%%%%%%%%%%%%%%%%%%%%%%%%%%%%%%%%%%%%%%%%%%%%%%%%%%%%%%%%%%%%%%%%%
{ \large{ {\bf {Appendix: Exact expression for the function  $ \Lambda(\mathcal{Z}^{I},\mathcal{P}^{I}) $}}}}\\
In the following we provide an exact expression for the function $ \Lambda(\mathcal{Z}^{I},\mathcal{P}^{I})  $ that appears in the expression for the Hamiltonian (\ref{E66}) beyond the pp wave limit.
\begin{eqnarray*}
\Lambda(\mathcal{Z}^{I},\mathcal{P}^{I}) =(1-\kappa^{2})(\sum_{i=1}^{4}(\mathcal{Z}^{i}_{R})^{2}\mathcal{P}^{2}_{i}+\frac{1}{2}\sum_{i\neq j}(\mathcal{Z}^{i}_{R}\mathcal{P}_{i})(\mathcal{Z}^{j}_{R}\mathcal{P}_{j}))+\kappa^{2}\sum_{m=5}^{8}(\mathcal{Z}^{m}_{\Theta})^{2}\mathcal{P}^{2}_{m}\nonumber\\
+\frac{1}{3}\sum_{m=5}^{8}\delta^{mn}((\mathcal{Z}^{k}_{\Theta})^{2}-\mathcal{Z}^{m}_{\Theta}\mathcal{Z}^{n}_{\Theta})\mathcal{P}_{m}\mathcal{P}_{n}
-\frac{1}{2}\left( \frac{1}{3}-\kappa^{2}\right) \sum_{m\neq n}(\mathcal{Z}^{m}_{\Theta}\mathcal{P}_{m})(\mathcal{Z}^{n}_{\Theta}\mathcal{P}_{n})
\nonumber\\
+2 \kappa\sum_{i,j,l=1}^{4}\sum_{a=1}^{2}\mathcal{P}_{i}(\mathcal{Z}^{l}_{R})^{2}(\mathcal{Z}^{j}_{R})^{2}\cos \Pi^{(R)} \sin \Pi^{(R)} \partial_{a}\Omega^{(R)} \partial_{i}\Pi^{(R)}\mathcal{Z}'^{a}_{R}\nonumber\\
-2 \kappa\sum_{m,n,p=5}^{8}\sum_{k=5}^{6}\mathcal{P}_{m}(\mathcal{Z}^{p}_{\Theta})^{2}(\mathcal{Z}^{n}_{\Theta})^{2}\cos \Pi^{(\Theta)} \sin \Pi^{(\Theta)} \partial_{k}\Omega^{(\Theta)} \partial_{m}\Pi^{(\Theta)} \mathcal{Z}'^{k}_{\Theta}\nonumber\\
-(1-\kappa^{2})\sum_{i=1}^{4}(\mathcal{Z}^{i}_{R})^{2}(\mathcal{Z}'^{i}_{R})^{2}-(1-\kappa^{2})\sum_{i\neq j}(\mathcal{Z}_{iR}\mathcal{Z}'^{i}_{R})(\mathcal{Z}_{jR}\mathcal{Z}'^{j}_{R})\nonumber\\
-\frac{1}{3}\left(\sum_{k,m =5}^{8} (\mathcal{Z}^{k}_{\Theta})^{2}(\mathcal{Z}'^{m}_{\Theta})^{2}-\sum_{m=5}^{8}(\mathcal{Z}^{m}_{\Theta})^{2}(\mathcal{Z}'^{m}_{\Theta})^{2}\right) -\kappa^{2}\sum_{m=5}^{8}(\mathcal{Z}^{m}_{\Theta})^{2}(\mathcal{Z}'^{m}_{\Theta})^{2}\nonumber\\
+\left( \frac{1}{3}-\kappa^{2}\right) \sum_{m \neq n}(\mathcal{Z}_{m \Theta}\mathcal{Z}'^{m}_{\Theta})(\mathcal{Z}_{n \Theta}\mathcal{Z}'^{n}_{\Theta}).
\end{eqnarray*}

\end{document}